# Detection and characterization of wind-blown charged sand grains on Titan with the DraGMet/EFIELD experiment on Dragonfly

**Audrey Chatain [a,*,1], Alice Le Gall [a,b], Jean-Jacques Berthelier [a], Ralph D. Lorenz [c], Rafik Hassen-Khodja [a], Jean-Pierre Lebreton [d], Tom Joly-Jehenne [e], and Grégoire Déprez [f].**

[a]LATMOS, Université Paris-Saclay, UVSQ, Sorbonne Université, CNRS, Guyancourt, France.

[b]Institut Universitaire de France (IUF), Paris, France.

[c]Johns Hopkins Applied Physics Laboratory, Laurel, MD, USA.

[d]LPC2E, CNRS, Université d'Orléans, Orléans, France.

[e]Université Paris-Saclay, ENS Paris-Saclay, 4 avenue des Sciences 91190 Gif-sur-Yvette, France.

[f]European Space Research and Technology Centre (ESTEC), Noordwijk, The Netherlands.

* Corresponding author: audrey.chatain@latmos.ipsl.fr

[1]Present affiliation: Departamento de Física Aplicada, Escuela de Ingeniería de Bilbao, Universidad del País Vasco/Euskal Herriko Unibertsitatea (UPV/EHU), Bilbao, Spain.

## Highlights

• EFIELD is a set of two electric field probes onboard the Dragonfly mission to Titan.

• EFIELD can detect wind-blown charged sand grains at Titan's surface.

• Insights on the charge and speed of grains can be retrieved from EFIELD measurements.

• An inversion approach is experimentally validated with a prototype EFIELD probe.

• An instrument noise sigma below 1 mV is required to study Titan wind-blown grains.

## Abstract

The EFIELD instrument is part of the geophysics and meteorology sensor package DraGMet on the Dragonfly mission, which will explore the surface of Titan in the mid-2030s. EFIELD consists of two electrodes designed to passively record the AC electric field at each Dragonfly landing site. The probes will be mounted at two different locations on the body of the rotorcraft lander.

The exploration zone of the Dragonfly mission will mostly consist of dune fields, covered with sand grains. Little is known on the properties of these grains, although Cassini-Huygens observations suggest they are mostly made of organic material produced by Titan's atmospheric photochemistry and evolved at the surface. Little is known also about dune formation and, in general, about the transport of sediments by winds on Titan. The latter much depends on inter-particle forces and therefore on whether grains are charged (e.g. by friction) or not. In this paper, we demonstrate that the EFIELD experiment can bring new insights on these questions.

We have developed a hydrodynamic-electrostatic model to simulate the trajectory of a wind-blown sand grain in the vicinity of an idealized EFIELD probe (an electronically insulated probe in a neutral open space) and to predict how such a grain flying close to the probe would affect its potential. We show that, in some conditions, the resulting perturbation will be strong enough to be detected by the EFIELD probe. More specifically, we find that the detection of typical charged wind-blown grains (200 μm in size) on Titan requires an instrument standard deviation noise inferior to 1 mV, though occasional larger grains flying close to one electrode could be detected with a higher noise level.

Furthermore, we propose a method to retrieve information on the charge and velocity of wind-blown charged grains detected by the EFIELD experiment. This method well applies for cases where the particle trajectory can be regarded as quasi-linear. We validate our inversion approach on both synthetic and experimental data obtained with a laboratory prototype of the EFIELD experiment. Such approach was developed with assumptions on the accommodation and electronic design of the EFIELD experiment but can be adapted to account for the final architecture of the experiment.

## Graphical abstract

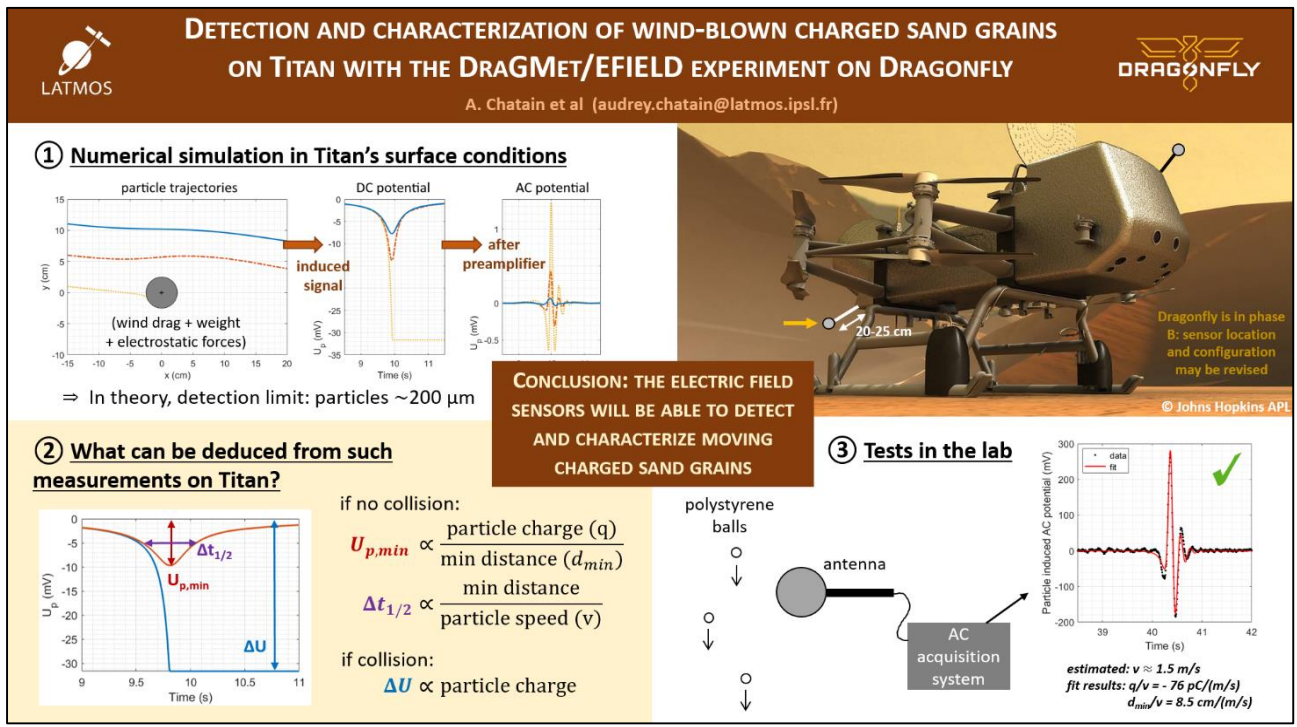

## 1- Introduction

Titan, Saturn's biggest moon, is an ocean world, whose surface is covered by organic materials and therefore one of the most promising astrobiological target in the Solar System, likely holding clues on the origin of life on Earth. NASA has selected the Dragonfly mission for its New Frontiers program in 2019. In 2027, the Dragonfly mission will thus send a rotorcraft lander to Titan (for an arrival in the mid-2030s) to investigate how far prebiotic chemistry has evolved on this moon, to characterize its environment and habitability, and to search for potential bio-signatures (Barnes et al., 2021; Lorenz et al., 2018; Turtle et al., 2018). Through multiple flights (20+ in 3.3 years), Dragonfly will explore a variety of locations and geologic settings, from the Shangri-La dune field to the rim of the young impact crater Selk (Lorenz et al., 2021).

To reach its scientific objectives, the Dragonfly rotorcraft is equipped with a sample acquisition system (DrACO, Zacny et al., 2019) and a payload of four instrument packages: a camera suite (DragonCam), a Gamma-Ray and Neutron Spectrometer (DraGNS, Parsons et al., 2018), a mass spectrometer with pyrolysis and gas chromatography capabilities (DraMS, Trainer et al., 2021) and, a Geophysics and Meteorology package (DraGMet). This latter includes a suite of sensors to monitor e.g. the temperature, pressure, methane abundance (humidity), wind speed and direction, ground dielectric constant, thermal properties and level of seismic activity at each investigated site. As presently anticipated (the DraGMet development is still in its preliminary design phase) it also includes two electric-field probes, referred to as the EFIELD experiment. This work focuses on the EFIELD experiment. The two EFIELD probes are designed to passively record the AC electric field at low frequencies (~5-100 Hz). They will be accommodated at two different locations on the rotorcraft, at the end of insulated booms protruding ~20 cm away from its main body, and pointing towards the exterior. Their design and configuration are still under discussion, but in this study, we consider conductive metallic spheres of 5 cm in diameter, electronically insulated and in a neutral open space.

The main objective of the EFIELD experiment is to measure Schumann Resonances (SR) on Titan. SR are ELF (Extremely Low Frequency) electromagnetic waves resonating between two electrically conductive layers. On Earth, they are triggered by lightning discharges and propagate in the resonant cavity formed by the ionosphere and the surface. On Titan, the ground is not conductive enough to serve as a waveguide but an underground salty ocean would. Therefore, measuring and monitoring SR on Titan would give insights into the depth of its buried ocean (Simões et al., 2008, 2007) and, more generally, into its habitability. EFIELD is a heritage of the Permittivity, Waves and Altimetry (PWA) analyzer, part of the Huygens Atmospheric Structure Instrument (HASI) which, during Huygens descent in Titan's atmosphere in 2005, has observed a possible SR at a frequency near of ~36 Hz (Béghin et al., 2012, 2009, 2007). This signal was interpreted to be the result of Saturn's magnetosphere exciting Titan's SR, with the ionized atmospheric layer (60-70 km altitude) serving as the top of the resonant cavity and a subsurface layer (55-80 km depth), presumably salty liquid water, as the lower boundary. However, this detection is subject to caution; it has been recently advanced that the detected signal could be due to a mechanical vibration during Huygens' parachute descent (Lorenz and Le Gall, 2020). The main goal of the EFIELD experiment is thus to perform electric field measurements from the surface over an extended period of time, with stable booms and without the mechanical noise of parachute descent to inform the existence of SR on Titan. If confirmed, the SR measured with a much better spectral resolution than PWA/HASI may offer constraints on the depth of Titan's buried ocean.

EFIELD has a secondary science goal, which is to detect and characterize near-surface wind-blown charged grains, if any. Winds on Titan can indeed put particles in saltation or suspension near the surface (Burr et al., 2015; Lorenz, 2014). In this process, the particles could get charged by friction (triboelectric charging; Méndez Harper et al., 2017) and would then induce a short transient perturbation on the potential of the EFIELD probes. It is well-known that blowing sand, dust and snow can generate transient signals on electrodes and these signals tend to quantitatively correlate with particle flux (e.g. Franzese et al., 2018). Indeed, a regression line fitted to a scatterplot of electric field fluctuation vs wind speed (Gordon and Taylor, 2009) has a zero-field intercept which is a good estimate of the saltation threshold. Measuring such perturbations may also provide valuable insights into the particle charge and velocity, as demonstrated on Earth's deserts (Seran et al., 2013). These particle properties are key to better understand the processes of dune formation and evolution and, more generally, sediment transport on Titan (Lorenz et al., 2018; Melosh, 2011). In particular, a better knowledge of the cohesive forces exerted on sand grains and of the wind shear velocity required to move them should help determine whether Titan's dunes were formed under current or past climate conditions (Diniega et al., 2017; Lucas et al., 2014; Mitchell and Lora, 2016; Tokano, 2010).

This work aims at investigating the ability of the EFIELD probes to detect and characterize individual sand grains blown by winds and charged by friction at the surface of Titan. For this purpose, we have developed a combined hydrodynamic and electrostatic numerical model to simulate the trajectory of a charged particle in the vicinity of an EFIELD electrode (section 2). We then show that the electrostatic influence or impact by a charged particle on an EFIELD probe will generate a signal strong enough to provide information on the charge and velocity of the wind-blown grains, especially in the cases where the particle trajectory can be regarded as quasi-linear (section 3). Lastly, our model and inversion approach are validated through laboratory experiments performed with a preliminary prototype of the EFIELD electrodes (section 4).

## 2- Modeling of the effect of a wind-blown charged particle on EFIELD measurements

In this section, we present the numerical model developed to simulate the detection of a single charged sand particle by one EFIELD electrode on Titan. The trajectory of the particle near the EFIELD probe is governed by three forces: gravity, the wind drag and the electrostatic force induced by the presence of the conductive EFIELD probe. When passing near the electrode, the particle induces a time-varying potential that, depending on the size and charge of the particle and its minimum distance to the electrode, should be detectable by EFIELD.

## 2.1- Hypotheses on the properties of Titan's sand particles

Little is known about the size and properties of Titan's sand grains, though Cassini-Huygens observations suggest they are organic in nature (e.g. Janssen et al., 2016; Lorenz et al., 2006). They are likely composed of organic materials, initially formed in the atmosphere by photo-chemistry before being deposited onto the surface where they accumulated over time resulting in a sedimentary layer then evolved by erosional processes such as river channel incision and aeolian abrasion (Brossier et al., 2018; Burr et al., 2013; Lapôtre et al., 2022). Based on laboratory measurements on Titan's aerosols analogues, we set the density of sand particles to a constant $\rho_p$ = 210 kg/m$^3$ (Brouet et al., 2016).

Furthermore, little is known about the wind speed required to lift Titan's particles off the ground, either for a very short moment (by saltation) or for a longer period of time (longer than a few seconds, by suspension). The size distribution of the particles that could be sensed by EFIELD is intimately related to the shear velocity threshold for wind-driven transport of grains (as a function of their size) on Titan. Particles in saltation are typically larger than suspended particles, but remain relatively close to the ground (e.g. Lorenz and Zimbelman, 2014, indicate a typical saltation height of 1cm). As a consequence, they could be detected only if the EFIELD probe is less than ~10 centimeters above the ground. In contrast, particles put in suspension should easily reach heights of the order of 1 m or more above the ground, but may generate smaller electric signals on the probe than saltated particles as they should, in average, be smaller. Nevertheless, it is possible that, under certain wind conditions that may be met on Titan (see below), sand-sized grains (50 μm to 2 mm in diameter) are put in suspension and therefore transported toward the EFIELD probes, even if these latter are not located close to the ground.

There is an optimum particle size for which the threshold wind speed for wind-driven transport is minimum. To assess this optimum size, it is key to understand the ability of Titan's particles to charge (Lorenz, 2022). In particular, triboelectric charging of sand grains could create strong inter-particle forces requiring higher wind speeds for saltation and suspension (Lorenz, 2014). This effect is likely enhanced on Titan because of its high atmospheric pressure (enabling high charge densities), the high bulk resistivity of the hydrocarbon material (on which charges can thus stay for a long time), especially in a cold and dry environment, the low conductivity near the surface (which slows down charge leakage), and the low density of the hydrocarbon grains (which makes them more sensitive to electrostatic forces). Laboratory experiments suggest that sand grains on Titan could have a grain charge-to-mass ratio of -10 to -100 μC/kg (Méndez Harper et al., 2017), values which were thus used in our model.

Wind tunnel experiments, partly accounting for triboelectric charging, led to an estimate of the optimum size for sand particles on Titan of ~250 μm in diameter (Burr et al., 2015); such particles should be put in saltation for a minimum friction speed threshold of 0.05 m/s. On Titan, this friction speed corresponds to a freestream air speed ~25 times larger just above the surface (Lorenz, 2014; Lorenz and Zimbelman, 2014). These experimental results may be compared to the very recent theoretical study by Comola et al. (2022) which takes cohesive forces into account. Comola et al. (2022) distinguish two mechanisms to lift particles and suggest that the optimum particle size and the associated threshold friction speed are different if the particle is directly moved by aerodynamic lifting ("fluid threshold"), or by the impact of a larger particle ("impact threshold"), also called the "granular splash" process. Their results give a minimum fluid friction threshold of 0.12 m/s for 2 mm particles, and a minimum impact threshold of 0.03 m/s for 100 μm particles. Consequently, the size distribution of particles lifted by winds dramatically change depending on the dominant regime (aerodynamic lifting or granular splash), which is dictated by the wind speed. In the same study, the authors implemented their sand lifting model in a general circulation model and they conclude that under Titan's conditions dunes are likely to be formed by the finest grain fraction. We therefore chose to investigate the trajectory of particles with radius ranging from 20 to 300 μm. Table 1 summarizes our assumptions on the particle properties.

| Parameter | Input values | References |
|---|---|---|
| Radius $r_p$ | 20 - 300 μm | (Burr et al., 2015; Comola et al., 2022) |
| Density $\rho_p$ | 210 kg/m³ | (Brouet et al., 2016) |
| Charge-to-mass ratio $q_m$ | -10/-100 μC/kg | (Méndez Harper et al., 2017) |
| Atmospheric density $\rho_{fl}$ | 5.3 kg.m⁻³ | (Fulchignoni et al., 2005; Lorenz, 2020) |
| Atmospheric viscosity $\mu_{fl}$ | 6.7 μPa.s | (Lorenz, 2020) |
| Initial wind speed $V_{fl,0}$ | 0.05 − 4 m/s | (Lorenz, 2021a) |
| Titan acceleration of gravity $g$ | 1.352 m.s⁻² | (Lindal et al., 1983) |

The importance of knowing the threshold wind speed for saltation (e.g. to interpret dune morphology to constrain Titan's paleoclimatology, Lorenz 2022) is such that Dragonfly will perform dedicated experiments on driving sand with the downwash (up to ~5 m/s) from its rotors when on the ground (Lorenz, 2017). It had originally been envisioned that sediment transport would be detected by before/after imaging, and, in real time, by electric field fluctuations. However, the operation of the rotors may be accompanied by strong electrical noise, and so as insurance against weak charging and/or heavy noise, photodiode detectors have been incorporated into the vehicle skids to also detect shadows of blowing particles. Electrical noise should be low, however, when the rotors are not operating and sediment is driven by ambient winds.

## 2.2- Wind drag and grain weight

### 2.2.1- Wind velocity around the EFIELD electrode

Lorenz (2021) investigates in details the wind expectations at the Dragonfly initial landing site and reaches the following conclusions: winds are very weak on Titan compared to those on Earth; the superposition of global circulation winds (< 1 m/s), local terrain induced flows (< 1 m/s) and occasional vortices (< 2.8 m/s) leads to a maximum wind of 4.8 m/s at 10 m height, corresponding to ~4 m/s at 1 m height due to surface friction. As a consequence, the wind specification values at the surface for the Dragonfly mission are: a minimum wind speed of 0.05 m/s, an average wind speed of 0.5-1.0 m/s and a maximum wind speed of 4 m/s. These values were taken as model values in our numerical simulations as shown in Table 1, which also indicates the assumed atmospheric density $\rho_{fl}$ and viscosity $\mu_{fl}$.

In order to assess the nature of the flow pattern (of initial velocity $V_{fl,0}$) around a EFIELD electrode (of radius $R_s = 2.5$ cm), we calculated the Reynolds number as follows:

$$R_e = \frac{2R_s \cdot V_{fl,0} \cdot \rho_{fl}}{\mu_{fl}} \qquad [2.1]$$

with $V_{fl,0}$ varying between 0.05 and 4 m/s, the Reynolds number varies between 2 000 and 160 000 indicating a turbulent flow around the electrode, with a possible transition regime between laminar and turbulent for 0.05 m/s < $V_{fl,0}$ < 0.075 m/s (that is 2 000 < $R_e$ < 3 000). On the contrary, the flow pattern around a sand particles of size $2r_p$ is laminar (see section 2.2.2) which implies that sand grains follow the wind streamlines.

Because there is no analytic solution of Navier-Stokes equations in the case of a turbulent flux, these equations have to be solved numerically, searching for a stabilized regime, in order to predict the spatial distribution of the wind velocity around the EFIELD electrode. For this purpose, we used the SimFlow Computational Fluid Dynamics simulation software (sim-flow.com), and selected an OpenFOAM steady-state solver for turbulent flows of compressible fluids (see more details in section S1 of Supporting Information). We defined a mesh with a maximum cell size of 8 mm far from the probe, with several layers of refinement around the probe (such that the minimum cell size in the boundary layer is 25 μm, see below) and in the flow tail. With the 5-cm electrode at the center of the simulation volume, the total size of the numerical box in centimeters is: -15 < x < 40 (horizontal) and -20 < y < 20 (vertical).

Simulations show that, after some time, a periodic stabilized regime is reached with 3 distinct regions: i) upstream the sphere and, more generally, away from the wake behind the sphere, the wind flow is constant with time, ii) just behind the sphere, there is no or only a weak wind flow, iii) at a given distance behind the sphere (10 cm for a wind velocity of 1 m/s), oscillating instabilities appear (see Figure 1b). However, at such distance, the electrostatic influence of a charged particle on the probe potential is small enough so that a slight change of the particle trajectory has no detectable effect. We therefore hereafter use for our calculations of trajectories the wind velocity map obtained as the oscillating instabilities are only incipient (Figure 1a).

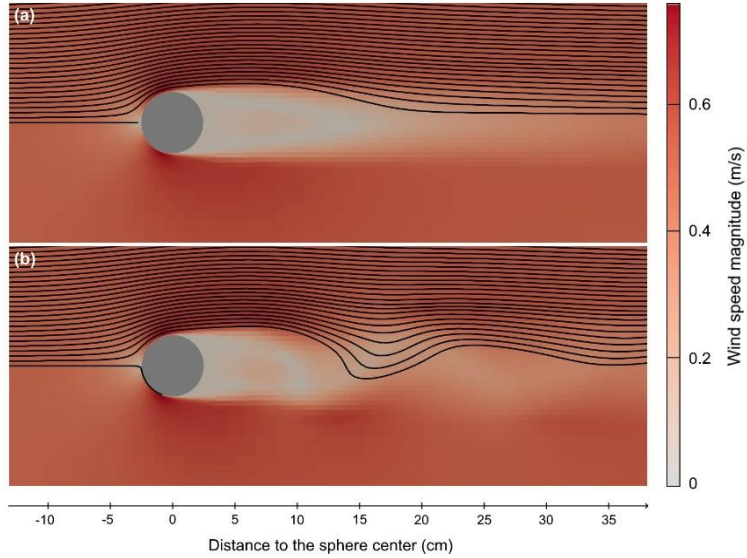

**Figure 1:** 2D maps of the wind speed magnitude around an EFIELD electrode obtained at two different time periods of the SimFlow simulation: (a) just before the beginning of the oscillating instability, and (b) during the oscillation. Streamlines are indicated with black lines on half of the panels for sake of clarity. The injected wind speed is here $V_{fl,0} = 0.5 \ m/s$.

As reported in Table 1, we investigated $V_{fl,0}$ ranging from 0.05 to 4 m/s. The boundary layer (i.e. the fluid zone close to the EFIELD electrode whose speed is strongly decreased by friction compared to the further free fluid) varies with $V_{fl,0}$: its width is of the order of $\delta = 2R_s/\sqrt{R_e}$ that is 1.1 mm for $V_{fl,0}$ = 0.05 m/s, 0.35 mm for $V_{fl,0}$ = 0.5 m/s and 0.13 mm for $V_{fl,0}$ = 4 m/s. We therefore set the cell size to 25 µm in the boundary layer, small enough to insure a proper spatial resolution of our model in this area for all tested wind conditions (see more in Supporting Information S1).

### 2.2.2- Wind drag on the particles

The general expression for the wind drag force $\overrightarrow{F_{drag}}$ exerted on a spherical particle of radius $r_p$ is given by:

$$\overrightarrow{F_{drag}} = -\frac{1}{2}\rho_{fl}.\left\|\overrightarrow{v_p} - \overrightarrow{V_{fl}}\right\|.\left(\overrightarrow{v_p} - \overrightarrow{V_{fl}}\right).\pi r_p^2.C_x \qquad [2.2]$$

where $\overrightarrow{V_{fl}}$ is the wind speed at the position of the particle, $\overrightarrow{v_p}$ is the particle velocity and $C_x$ is the drag coefficient for a sphere. This latter depends on the Reynolds number associated to the flow around the particle, namely $R_e = \rho_{fl}.\left\|\overrightarrow{v_p} - \overrightarrow{V_{fl}}\right\|.2r_p/\mu_{fl}$. It can be estimated from the analytical-empirical curve provided by Clift et al. (1978), which gives the variations of $C_x$ as a function of $R_e$ (see section S2 in Supporting Information). SimFlow-derived 2D velocity maps are used for $\overrightarrow{V_{fl}}$. The initial wind direction is assumed horizontal (which is generally the case over flat surfaces). We assumed spherical particles in this work by lack of more information on Titan's sand grains and by comparison to the Earth where sand grains are mostly spherical. Dragonfly's images of the surface at the microscopic scale (Barnes et al., 2021) will ultimately infirm or confirm this assumption.

In addition to the wind drag and, not yet accounting for electrostatic effects, the particle is subjected to its own weight and the Archimedes' buoyant force (which is however negligible compared to the wind drag and gravity). These can be expressed as follows:

$$\overrightarrow{F_{w+A}} = \frac{4}{3}\pi r_p^3.(\rho_p - \rho_{fl}).\vec{g} \qquad [2.3]$$

Before its arrival in the vicinity of the EFIELD electrode, the particle should be at the equilibrium between the wind drag and its weight. Therefore, the trajectory of the particle far from the sphere is a line, forming an angle $\alpha$ with the horizontal direction. For our study, particles are injected at 5 m from the sphere along this linear trajectory, with the equilibrium speed as initial speed (see computation in section S3 in Supporting Information):

$$\begin{cases} v_{p,x} = V_{fl,0} \\ v_{p,y} = -\sqrt{\dfrac{8}{3C_x}.r_p.g.\dfrac{\rho_p}{\rho_{fl}}} \end{cases} \qquad [2.4]$$

### 2.3- Electrostatic forces

The EFIELD probe is represented by a conductive sphere of radius $R_s = 2.5$ cm electrically insulated. In reality, the probe is connected to a high impedance amplifier for measurements, whose effect on the probe potential is negligible (see details in section 4.2). We also consider that the probe is in an infinite open space environment (this hypothesis is discussed later in section 4.2) which is assumed to be electrically neutral, thus similar to vacuum. Measurements by the Huygens lander indeed concluded on the very low conductivity at Titan's surface (Hamelin et al., 2007; Lorenz, 2021b). However, the presence of a Multi-Mission Radioisotope Thermoelectric Generator (MMRTG[1]) onboard Dragonfly may locally enhance the atmospheric conductivity (Lorenz and Clarke, 2020). As discussed in (Lorenz, 2021b), the resulting value of the conductivity created by a RTG in Titan's conditions cannot be predicted with certainty, but the terrestrial atmosphere measured near the MMRTG should be somewhat representative. There, an insulated conductor had a relaxation time during which most of its charge leaked away within about 1 minute (compared to ~1 hour in the unperturbed atmosphere). Since this timescale is still large compared with the time to advect a sand grain past the lander (~1 second), the atmosphere can still be considered to be insulating.

#### 2.3.1- Charge by influence and electrostatic attraction by the EFIELD probe

A charged particle close to the probe induces a charge distribution on its surface which, in turn, generates an electrostatic attractive force on the particle. As developed in section S4 of the Supporting Information, this force can be expressed as:

$$\overrightarrow{F_{el1}} = -\frac{1}{4\pi\epsilon_0}.\frac{q_p^2.R_s.D}{(D^2 - R_s^2)^2}\overrightarrow{e_r} + \frac{1}{4\pi\epsilon_0}.\frac{q_p^2.R_s}{D^3}\overrightarrow{e_r} \qquad [2.5]$$

with $D$ the distance of the particle P to the center of the sphere O, $\overrightarrow{e_r} = \overrightarrow{OP}/\|\overrightarrow{OP}\|$, $q_p$ the particule charge ($q_p = q_m\rho_p\frac{4}{3}\pi r_p^3$) and $\epsilon_0$ the vacuum permittivity.

#### 2.3.2- Repulsion by the charged probe

Particles tend to charge negatively by friction (see section 2.1), those hitting the probe transfer their charges to it, and the probe therefore gets negatively charged with a charge $Q_s$. This results in an electrostatic repelling force exerted by the probe on the charged particles traveling in its vicinity namely:

---

[1] Dragonfly has been designed to use the same MMRTG as used on the Curiosity and Perseverance rovers. The use of a radioisotope power system on Dragonfly in subject to approval per the National Environmental Policy Act (NEPA) and thus reference to the MMRTG should be considered pre-decisional.

$$\overrightarrow{F_{el2}} = -\frac{1}{4\pi\epsilon_0} \frac{q_p \cdot Q_s}{D^2} \cdot \overrightarrow{e_r} \qquad [2.6]$$

It is worth pointing out that on Titan the sedimentation speed of sand grains once they are in suspension is much smaller than on Earth because of the lower gravity and the higher atmospheric pressure. The resulting high efficiency of particles blown by winds to charge Dragonfly and the EFIELD probes is thus an adverse effect that may lead to large negative potentials up to values corresponding to the runaway breakdown threshold electric field strength estimated to be ~200 kV/m near Titan's surface. For this reason, the vehicle body will be electrically conductive to prevent localized charge accumulation and dangerous electric breakdowns (Lorenz, 2020). In our numerical simulations, we therefore consider the possibility of a negatively charged electrode (up to -1 kV) with respect to a reference potential (which will on Titan be the lander body from which the probe is electrically insulated).

## 2.4- Particle trajectories

Figure 2 illustrates the geometry of the problem and the forces that are exerted on a charged particle passing near the EFIELD probe. The Newton's second law is numerically solved to compute the particle trajectory that is its position as a function of time. An implicit Euler scheme is used for discretization of Newton's second law which expresses as follows:

$$m_p \frac{d\overrightarrow{v_p}}{dt} = \overrightarrow{F_{drag}} + \overrightarrow{F_{w+A}} + \overrightarrow{F_{el1}} + \overrightarrow{F_{el2}} \qquad [2.7]$$

with $m_p$ the mass of the particle.

The time step $dt$ is adjusted depending on the wind speed, the mesh resolution and the particle weight in order to optimize the computation time and warrant the simulation stability. To better appreciate the combined effects of hydrodynamics, electrostatics and gravity, particles are initially positioned in the plane defined by the initial wind speed vector and the acceleration of gravity, passing through the center of the electrode. This way, all particles stay in this 2D plane during the simulation, making it easier to visualize.

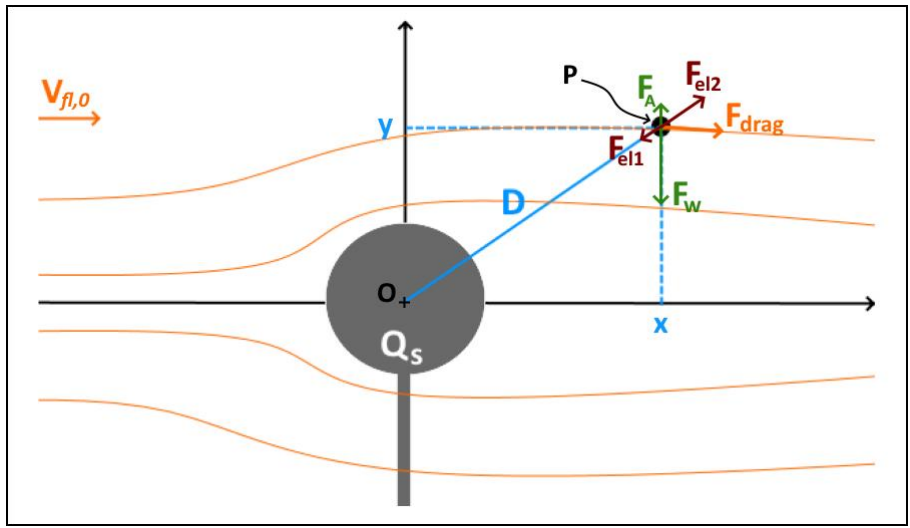

**Figure 2: Geometry and forces of the problem.**

The trajectory of a given particle depends on four parameters: the radius of the particle $r_p$, the initial wind speed $V_{fl,0}$, the potential of the electrode $U_s = \frac{Q_s}{4\pi\epsilon_0 R_s}$ and the particle charge-to-mass ratio $q_m$. Different cases are compared in Figure 3 which displays the trajectories of particles starting at different altitudes. We have considered that particles impacting the EFIELD electrode remain stuck to it.

The upper panel of Figure 3 (a,b,c) shows that stronger winds induce more horizontal trajectories due to a stronger $F_{drag}$. The middle panel (d,e,f) establishes that large particles tend to have rectilinear trajectories (i.e. less easily deviated) due to their high inertia. However, having a higher weight, they also tend to fall behind the electrode and cross the tail in which the wind is weaker. The trajectory deviation induced by the crossing of the tail is clearly visible in cases where the particle weight is larger than the wind drag (Figure 3a and 3f). As shown on the lower panel (g,h,i,), the charge of the electrode has not a strong effect on the particle trajectory, at least up to $U_s = -1$ kV for typical particles of 100 μm radius and charge-to-mass ratio of -100 μC/kg (i.e. with a total charge of -88 fC). We further note that the effect of $F_{el1}$ is negligible compared to the effects of wind drag and gravity. Studies (not shown here) on the effect of the electrostatic forces in the case of a smaller particle charge-to-mass ratio (typically $q_m = -10$ μV/kg), show that higher sphere charges (typically tenfold) are required to result in the same disturbance in the particle trajectories.

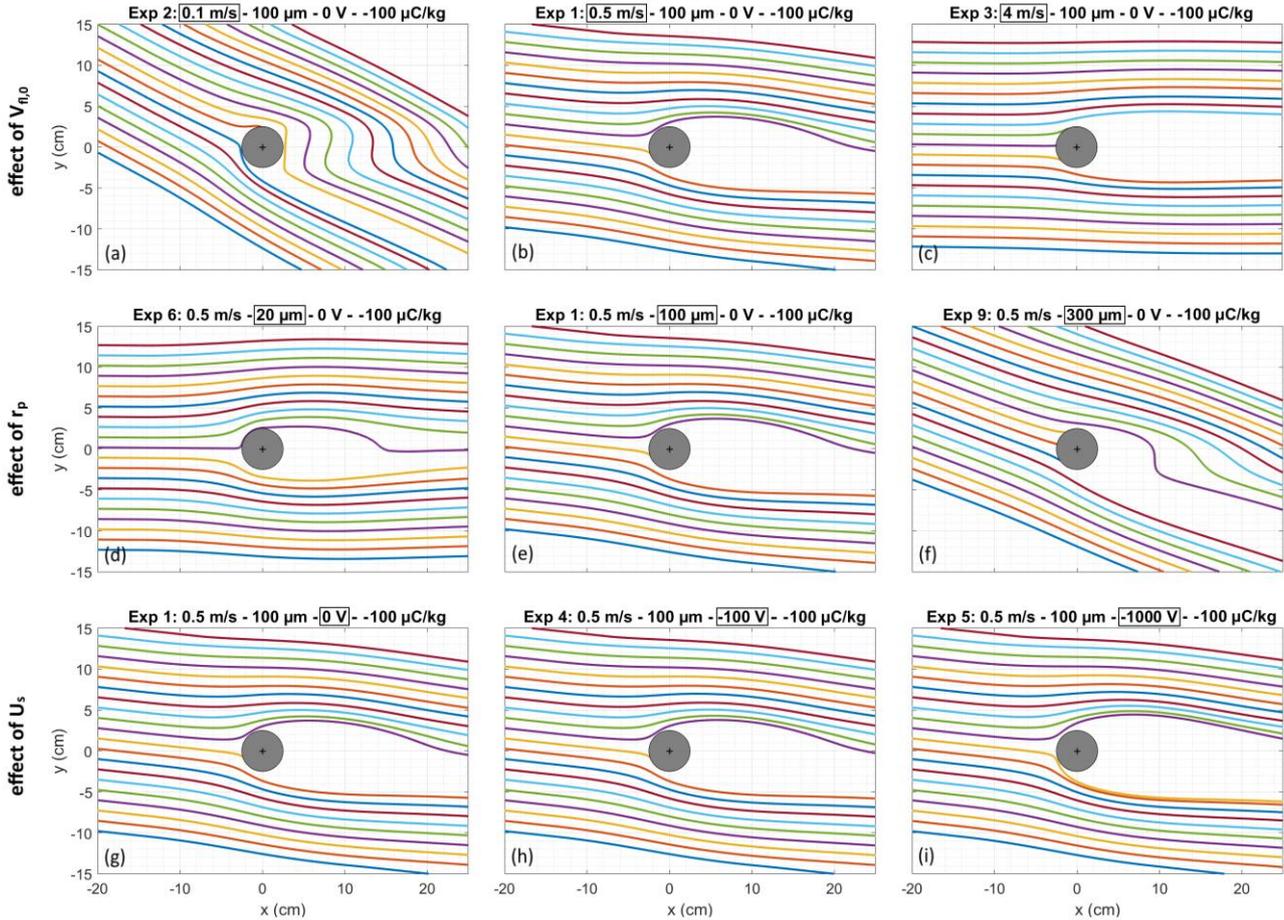

**Figure 3:** Simulations of the trajectory of particles (colored lines) under different conditions.

## 2.5- Signal induced on the potential recorded by a floating EFIELD probe

In the case where the particle collides with the EFIELD electrode and stays stuck on it, the particle transmits its charge and induces an offset $\Delta U$ on the DC potential of the sphere (Figure 4). The charge of the particle $q_p$ can then be directly retrieved from this offset with:

$$q_p = 4\pi\epsilon_0 . R_s . \Delta U \qquad [2.8]$$

Note that a particle entering the boundary layer of the flow around the electrode will most likely collide with the electrode.

In the more general case of a particle travelling close to the EFIELD probe without collision, the DC potential $U_p$ induced on the electrode by the charged particle varies with time $t$ according to the distance of the particle to the electrode $D(t)$ which can be computed with the model described in section 2.4 (Figure 4):

$$U_p(t) = \frac{1}{4\pi\epsilon_0} \cdot \left(\frac{4}{3}\pi\rho_p q_m \cdot r_p^3\right) \cdot \frac{1}{D(t)} \propto \frac{q_m \cdot r_p^3}{D(t)} \qquad [2.9]$$

We highlight that equation [2.9] is valid in the ideal case of a probe perfectly insulated from the ground potential, i.e. with a preamplifier with an infinite input impedance.

Equation [2.9] can be used to compute $U_p(t)$ (Figure 4) and also to estimate the theoretical possibility to detect particles from the induced DC potential variations of the probe. If the floating potential of the probe is of the order of a few volts one could expect that a threshold for the detection of the signal shown in Figure 4 could be ~1 mV. This corresponds to particles travelling within 1 cm (respectively 10 cm) of the probe surface if they have a radius larger than 30-40 μm (resp. 50-60 μm) in the case $q_m$ = -100 μC/kg. The minimum size increases with increasing distance to the sphere (see details in section S5 in Supporting Information).

However, estimates of the floating potential of the probe have shown that much larger voltages can be reached of the order of 10 to 100kV (Lorenz, 2020), leading to detection threshold of a few volts thus totally out of range. Therefore, we cannot rely on DC measurements and we should rather use AC acquisitions with the EFIELD instrument. The deformation of the signal from DC to AC results from the application of the transfer function of the AC pre-amplifier. In the frequency domain it simply corresponds to the multiplication of the DC signal by a frequency-dependent complex filter function $\widetilde{G(f)}$) as follows:

$$U_p^{AC}(t) = IFT(U_p^{DC}(f) \times \widetilde{G(f)}) \qquad [2.10]$$

where IFT stands for an Inverse Fourier Transform.

For a comparison with laboratory measurements (section 4), we use in our model the transfer function of the pre-amplifier of the EFIELD prototype electrode described in section 4 (the transfer function is displayed in section S-6 in Supporting Information) (Hamelin and Chabassière, 2011). Figure 4 illustrates the effect of three different particle trajectories on the time-varying potential recorded in both DC and AC modes.

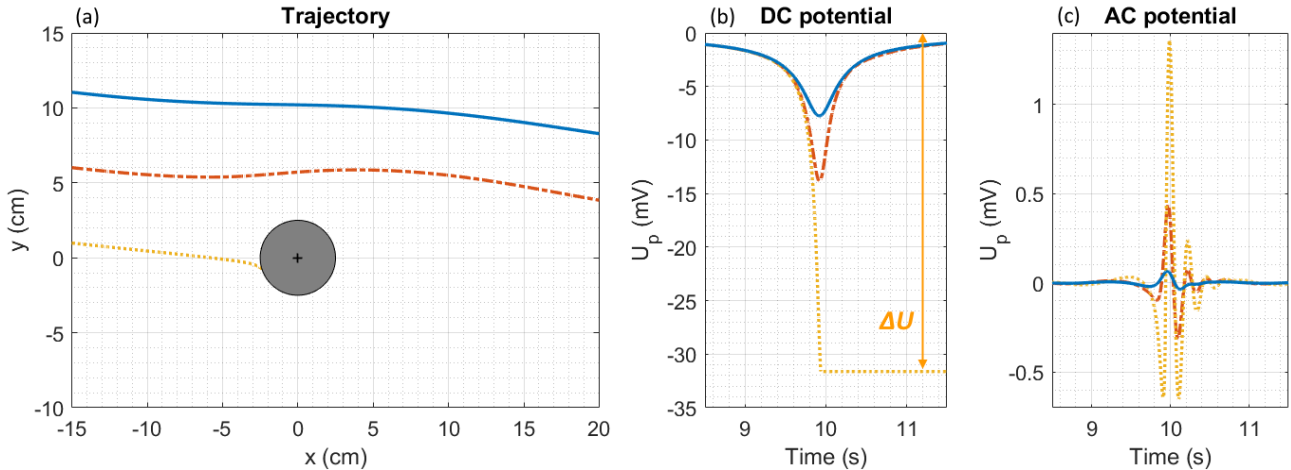

**Figure 4:** Simulated (a) trajectories and corresponding time-varying (b) DC and (c) AC potentials $U_p$ on a conductive sphere of $R_s$ = 2.5 cm induced independently by three negatively charged particles with different initial positions. All particles have a density charge $q_m$ of -100 μC/kg and a radius $r_p$ of 100 μm, and the wind velocity is $V_{fl,0}$ = 0.5 m/s. The conductive sphere is not charged here ($U_s$ = 0 V).

In conclusion, the model we developed to simulate the trajectory of a charged particle nearby EFIELD electrodes can be used to predict the temporal variations of the potential (equation [2.9]) that will be measured by DraGMet on Titan. With a carefully designed low-noise electronic circuit, charged sand-particles should be detectable by the EFIELD probes within reasonable ranges of size, charge-to-mass ratio and minimum distance to the probe. In turn, we will demonstrate (sections 3 and 4) that the variations of potential recorded by the EFIELD experiment can be used to retrieve some of the characteristics of the detected grains.

# 3- Retrieval of sand grain properties: validation on simulated signals

This section presents the inversion approach we have developed to retrieve information on the properties of a moving charged particle describing a quasi-linear trajectory from its induced signal on the EFIELD probe.

## 3.1- Simplified expression of the potential variations induced by a charged particle

The results obtained in section 2 show that most of the charged particles driven by wind in the vicinity of the EFIELD electrode follow a quasi-linear trajectory with a quasi-constant speed. For these particles, we can approximate their trajectory to a straight line and their speed to a constant $v_{p,0}$ (see Figure 5), and use the Coulomb law to readily deduce the induced potential variations with time from the particle trajectory (equation [3.2]). Such method is especially valid for heavy particles, strong wind conditions and/or particles far from the sphere (see Figure 3). Its conditions of application are tested further in this work.

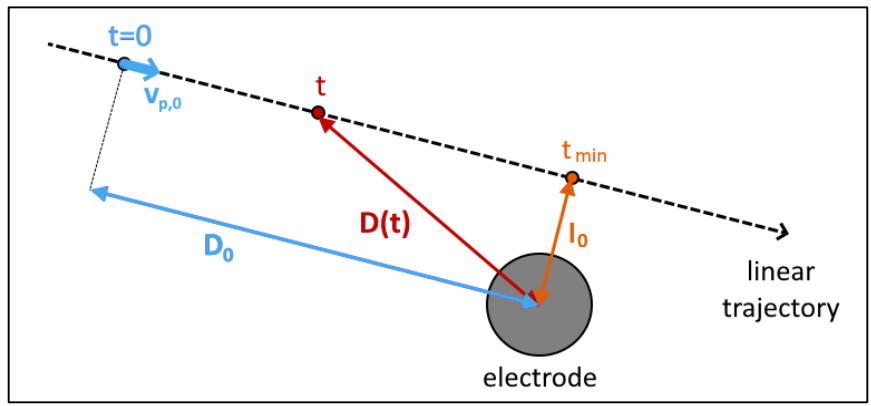

**Figure 5:** Notations used for the computation of the potential time variations.

The particle distance to the EFIELD electrode is expressed as:

$$D(t) = \sqrt{\left(v_{p,0} \cdot t - D_0\right)^2 + l_0^2}$$ [3.1]

with $l_0$ the impact parameter (that is the minimum distance between the particle and the electrode) and $\sqrt{D_0^2 + l_0^2}$ the distance to the electrode at $t = 0$.

The induced DC potential then follows:

$$U_p(t) = \frac{1}{4\pi\epsilon_0} \frac{q_p}{D(t)} = \frac{1}{4\pi\epsilon_0} \frac{q_p/v_{p,0}}{\sqrt{(t - t_{min})^2 + \left(l_0/v_{p,0}\right)^2}} \quad \text{with} \ \ t_{min} = D_0/v_{p,0}$$ [3.2]

The maximum amplitude of the induced DC potential is:

$$U_{p,max} = U_p(t = t_{min}) = \frac{1}{4\pi\epsilon_0} \frac{q_p}{l_0}$$ [3.3]

The full width at half maximum (FWHM) of the induced DC potential signal ($\Delta t_{1/2}$) is:

$$U_p = \frac{U_{p,max}}{2} \qquad \Leftrightarrow \frac{1}{4\pi\epsilon_0} \frac{q_p}{\sqrt{\left(v_{p,0} \cdot t - D_0\right)^2 + l_0^2}} = \frac{1}{4\pi\epsilon_0} \frac{q_p}{2l_0}$$

$$\Leftrightarrow \left(v_{p,0} \cdot t - D_0\right)^2 + l_0^2 = 4l_0^2$$

$$\Leftrightarrow t = \frac{\pm\sqrt{3}l_0 + D_0}{v_{p,0}}$$

$$\Rightarrow \Delta t_{1/2} = \frac{2\sqrt{3}.\, l_0}{v_{p,0}} \qquad [3.4]$$

According to this simplified model, the amplitude of the DC signal induced by a charge particle as it travels close to an EFIELD electrode is directly proportional to the charge of the particle, while the width of the perturbation is proportional to the inverse of the particle velocity. In the case of particles for which the linear trajectory hypothesis is valid, the equations above can thus be used to retrieve information on any detected particle. More specifically, fitting EFIELD measurements with equation [3.2] (after accounting for the AC pre-amplifier transfer function, see section 2.5) should give access to the values of the ratios $q_p/v_{p,0}$ and $l_0/v_{p,0}$. Further, since on Titan the velocity of the particles can also be estimated from wind measurements and cameras, the particle charge can be derived from the inverted $q_p/v_{p,0}$ value. This approach is tested in the following sections (where $v_{p,0}$ is simplified by $v_p$), first on simulated data (section 3.2) and then on experimental observations (section 4).

## 3.2- Validation of the inversion approach on simulated data

Figure 6 shows several simulated AC signals. The noise-free signal (in blue) is compared to more realistic signals on which a white Gaussian noise with a standard deviation $\sigma_n$ of 3 mV, 0.5 mV or 0.1 mV was added. Fits obtained with the simplified model previously described using a nonlinear least-squares optimization algorithm (trust region reflective) are also plotted, for the different noise cases. For bigger particles and/or particles closer to the electrode, the fit to the data is very satisfying in all tested noise conditions and the retrieved parameters are very close to the input ones. As expected, the results degrade as the particle gets further or smaller and as the noise increases. Nevertheless, typical 200 µm particles with a charge-to-mass ratio of -100 µC/kg (i.e. with a total charge of -88 fC) can easily be detected at 3.3 cm from the electrode center with a noise standard deviation of 0.5 mV. Figure 6i-j shows that the signal induced by a 600 µm particle (i.e. with a total charge of -2.4 pC) could even be observed with a $\sigma_n$= 3 mV noise level (which is the current noise level of the prototype electrode used in section 4). All fits and retrieved parameters corresponding to noise-free conditions are perfect, except in isolated cases where the trajectories of particles cannot be regarded as linear (see Figure 6g-h and typical deformed trajectories in Figure 3a). The good fits obtained attest that the approximations done in section 3.1 are generally valid for this type of analysis.

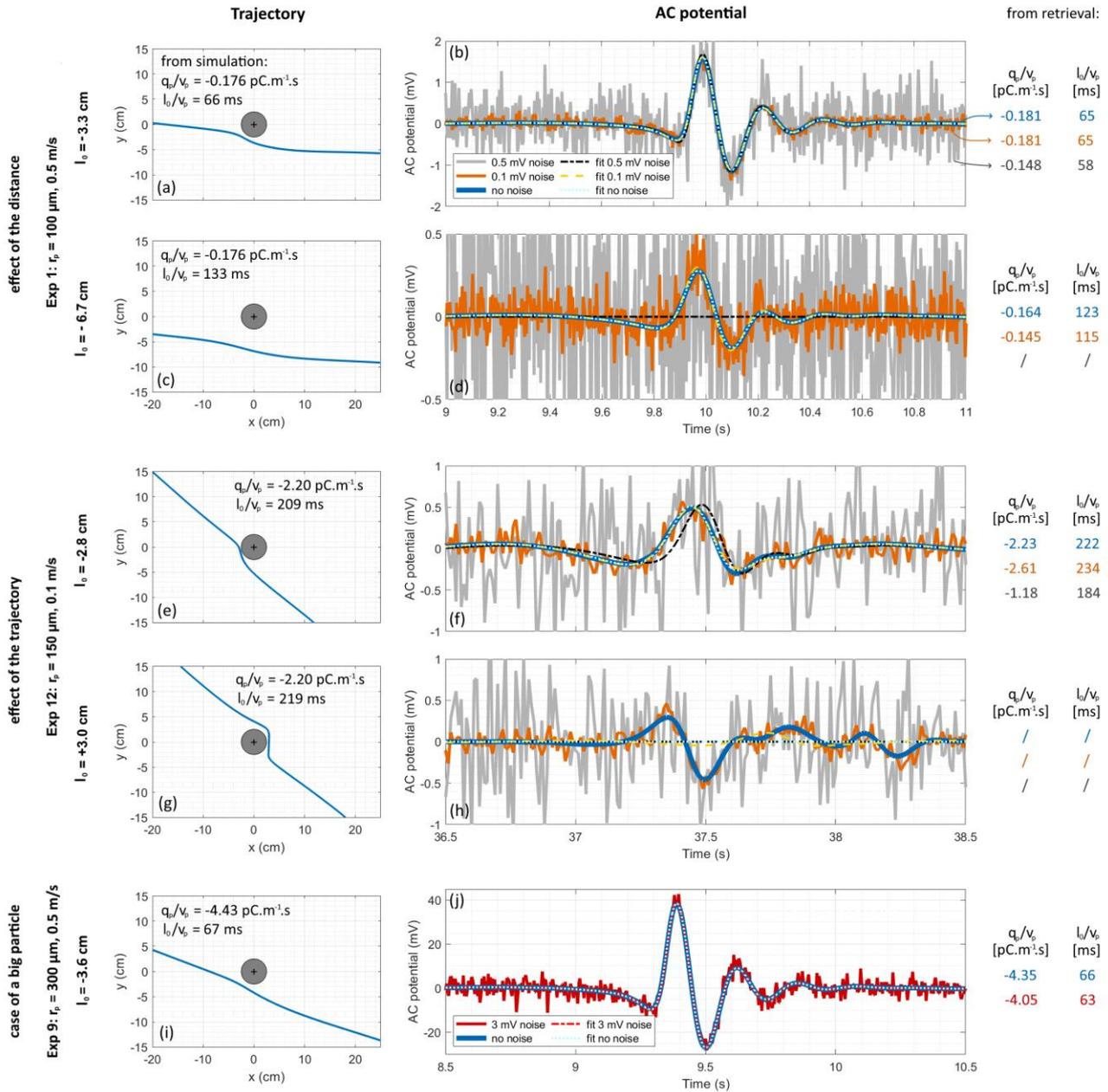

**Figure 6:** Examples of simulations of particle trajectory (left) and induced signal in AC (right) mode, without noise and with the addition of a white Gaussian noise with standard deviations equal to 0.1 mV, 0.5 mV and/or 3 mV. Fitting in all conditions are done with equation [3.2] and the transfer function plotted in Figure S6-1. All the plotted cases correspond to $U_s = 0$ V and $q_m = $ -100 µC/kg.

More detailed analysis is given in the following figures. Figure 7 investigates the robustness of the inversion approach in the conditions of experiment 1 ($r_p = 100$ µm, $V_{fl,0} = 0.5$ m/s, $U_s = 0$ V, $q_m = $ -100 µC/kg, see Figure 6abcd). 21 initial particle locations are simulated. The resulting synthetic signals are fitted with the method presented above. Several noise level cases are considered. The inversion results for $q_p/v_p$ and $l_0/v_p$ for all initial particle locations and noise level cases are compared to the values of these parameters given by the simulations. Figure 7 shows the relative error and confidence interval (at 95%) on the retrieved $q_p/v_p$ and $l_0/v_p$ values given by the fitting algorithm, as a function of noise level and minimum distance of the particle to the center of the electrode. For a robust statistical approach, we simulated each case of study 100 times with a random Gaussian noise. We then computed the average of the errors and the average of the best-fit confidence intervals over the 100 tests. A given fit is considered unsuccessful (ie. no value is given to the fitted parameters) if less than 50% of the tests have led to convergence.

- 13 -

All distances from the probe are larger than 2.5 cm, the probe radius. The farthest tested particle has a minimum distance of 12 cm in this case. All other regions of the graphic left blank correspond to non-detected particles or particles whose signal cannot be fitted, which happens for the highest noise levels, and especially for particles far from the electrode. In all cases where particles are detected and their signal fitted, the errors on the retrieved parameters are usually smaller than 15% for $l_0/v_p$. The error on $q_p/v_p$ can reach 50% to 80%, but these are still very good retrievals because particle charges vary over several order of magnitude. The confidence interval obtained is always larger than the error. This proves that outputs given by the simple fit method are reliable.

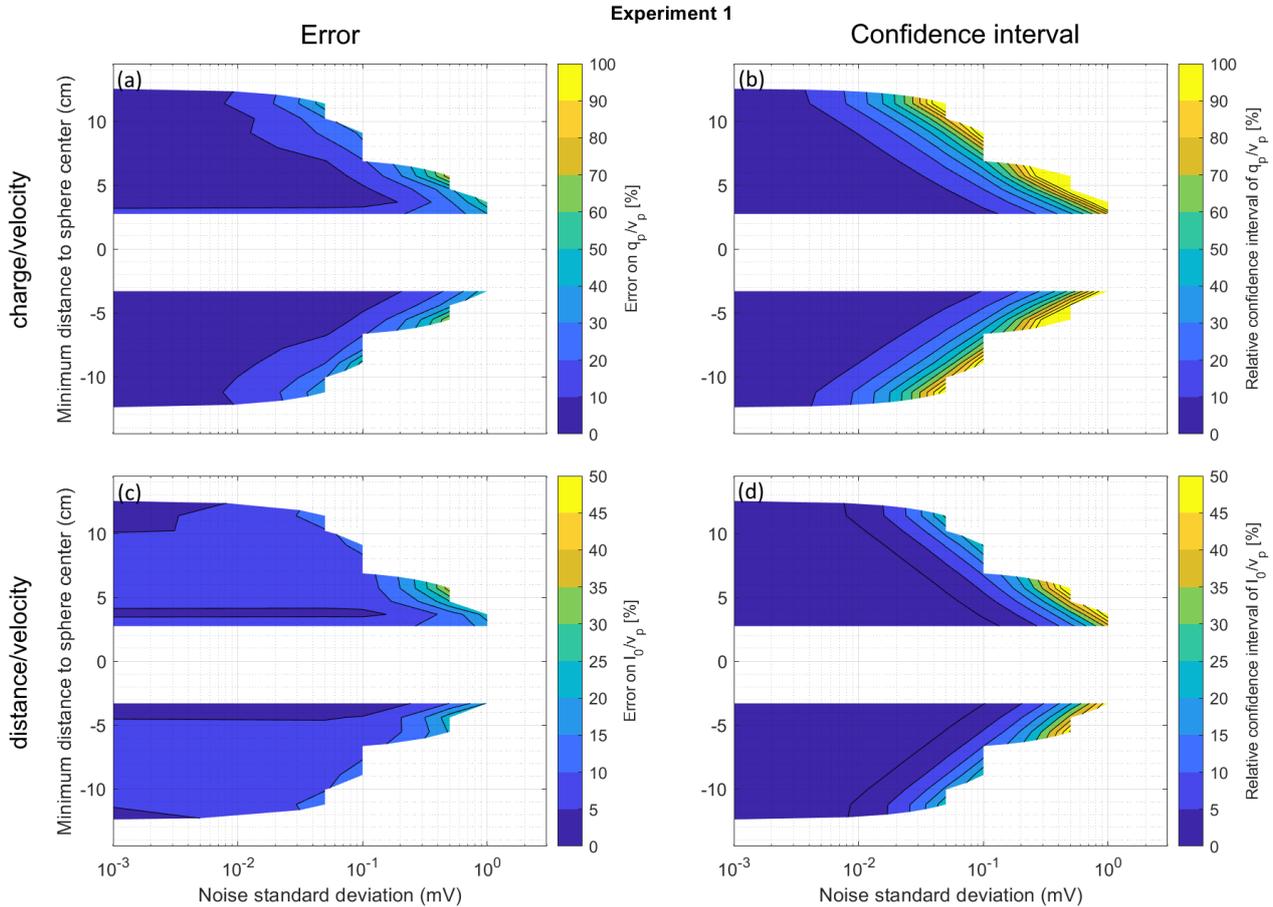

**Figure 7:** Error and confidence interval on the retrieved parameters as a function of the noise level and the particle distance to the electrode, in conditions of experiment 1 ($r_p$ = 100 μm, $V_{fl,0}$ = 0.5 m/s, $U_s$ = 0 V, $q_m$ = -100 μC/kg). Errors and confidence intervals are averaged over 100 tests.

Though not shown here, simulations obtained in a wide range of conditions were tested (see examples in section S7 in Supporting Information). In nearly all cases, the derived parameters obtained for noise-free cases are in perfect agreement with the values used in the simulations. The only exceptions are the isolated cases where the trajectories of particles are not linear and thus where equation [3.2] does not apply, like for the particles of experiment 12 falling behind the electrode in the turbulent tail (see Figure 3a). Similar as for experiment 1, we note that when particles are detected and their signal can be fitted, the parameters are retrieved with reasonable error (typically <15-20% for $l_0/v_p$, <30% for $q_p/v_p$, and often <5-10% for both parameters).

Because of the used AC transfer function, the speed of the particles impacts strongly the ability of the electrode to detect them: AC signals of the slowest particles have a lower maximum intensity than AC signals of the fastest particles. Consequently, it is more difficult to detect and characterize slowly-moving particles. Small (hence little charged) particles are also less easily detected than large ones. Figure 8 illustrates the errors on the retrieved parameters as a function of the particle size and noise level, for two different wind conditions.

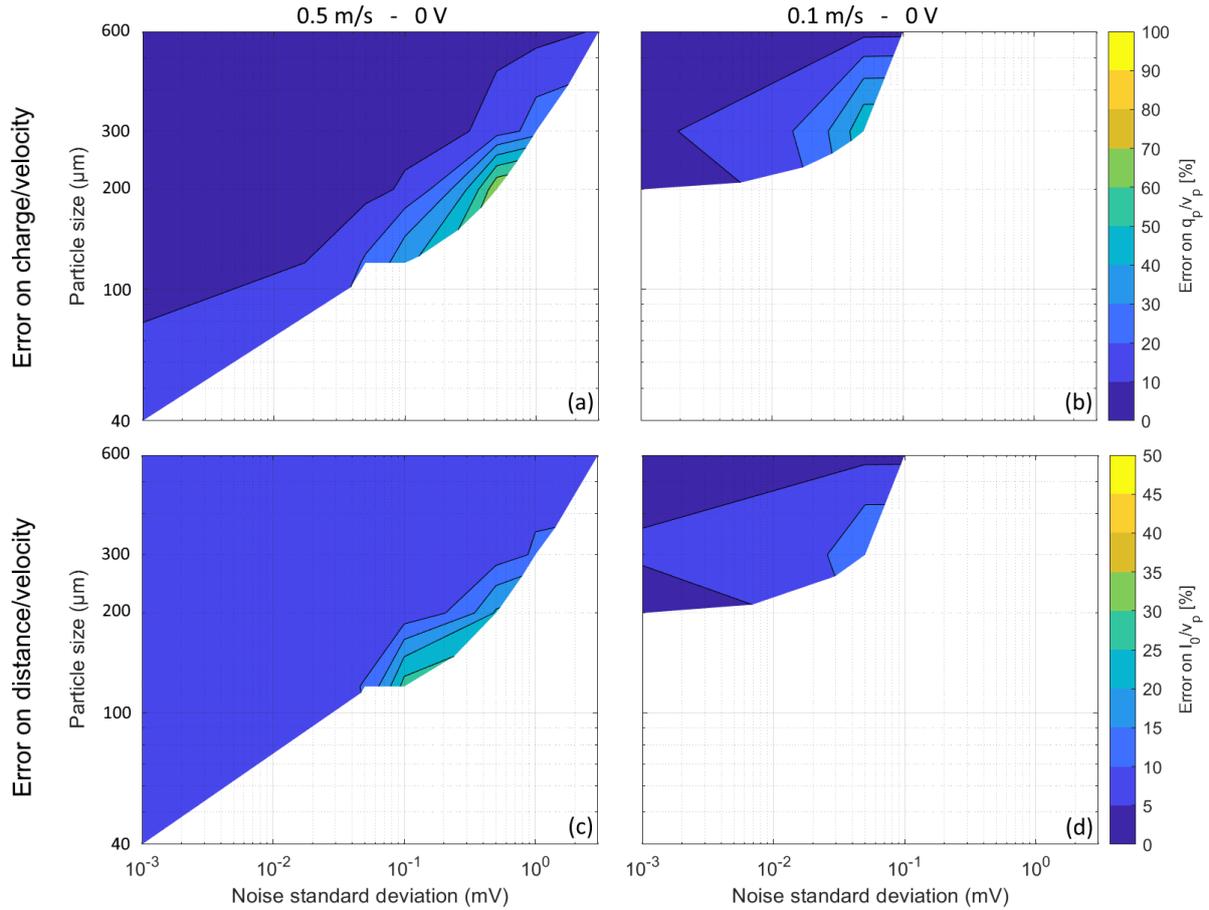

We have shown that as long as a particle can be detected and the AC signal it induces can be fitted (ie. when the fitting algorithm converges) the input parameters are retrieved with a reasonable error. A critical point is therefore to be able to detect and to fit the particle signals. In Figure 9, we investigate the maximum noise level for which particles can be detected and their signals fitted, in various simulation conditions. Typical 200 μm-particles driven by 0.5-m/s winds give clear, analyzable signals up to distances of 7 cm from the electrode center with a $\sigma_n$ of 0.1 mV, and up to 5 cm with $\sigma_n$= 0.5 mV. In contrast, particles of 40 μm cannot be detected for the tested noise levels ($\sigma_n$= 0.05, 0.1, 0.5, 1 and 3 mV). The detection and characterization of 600 μm particles is guaranteed for $\sigma_n < 1$ mV. The charging of the electrode up to -1000 V does not impact much the results.

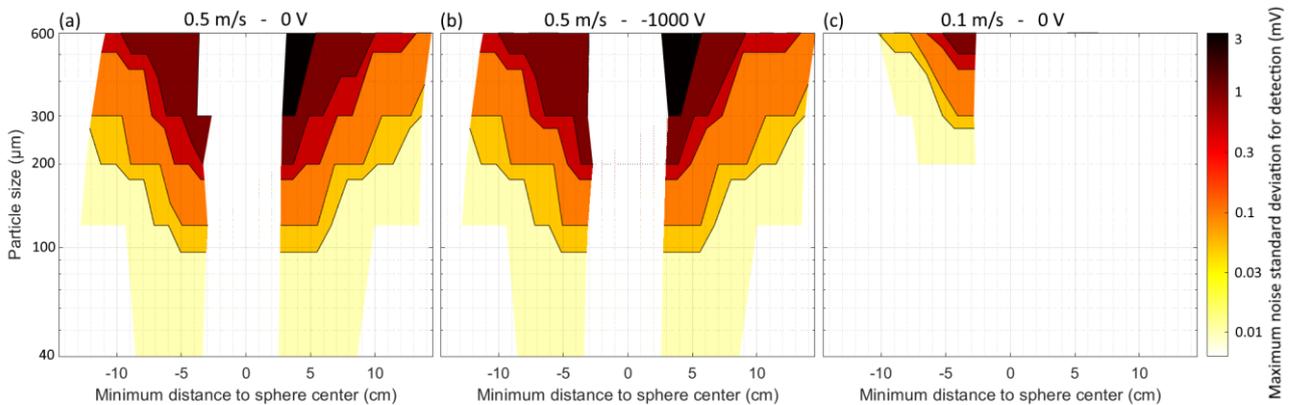

**Figure 9:** Threshold noise level to detect and fit a charged particle signal, as a function of the particle size and its minimum distance to the electrode center. Results are averaged over 100 tests.

# 4- Experimental validation with a prototype electrode

To further validate our model and the inversion approach described in the previous section, we have conducted experimental tests in an electrostatically controlled environment with a preliminary prototype of the EFIELD electrode and analogues of Titan's charged sand grains. The results are very promising and point to some technical requirements and measurement specifications for the EFIELD experiment.

## 4.1- Experimental setup

To build a prototype of the EFIELD electrode we used a EM (Engineering Model) of the Huygens/PWA (Permittivity, Wave and Altimetry) experiment (Fulchignoni et al., 2002) by replacing the original 10 cm diameter ring by a spherical aluminium probe with the same diameter. The electronics for the AC (RX) channel are housed in a small box close to the short boom supporting the probe (Cadène, 1995) and supplied by two 12 V lead batteries to minimize any external electromagnetic noise due to power supplies. It must be noted that the probe radius is twice the value used in the simulation and planned for Dragonfly but the computations were adapted to this value. The signals are observed and recorded through a numerical oscilloscope (Analog Discovery from Digilent).

Charged sand particles are experimentally simulated by small polystyrene balls charged in a rotating cylinder, itself coated with small polystyrene balls. This technique, which aims at simulating the grain charging by friction induced by winds at the surface of Titan, has been previously used and characterized in Méndez Harper et al. (2017). Balls are dropped one by one from ~20-25 cm above the electrode to fall at a maximum distance of 3-4 cm from the electrode's surface. Balls of 5 mm, 3 mm and 0.2 mm in diameter have been tested, though it turned out that the sensitivity of the system was too limited to detect the 0.2 mm balls. The current pre-amplifier has a noise of ~3-4 mV standard deviation (note: the bandwidth is ~3 to 6000 Hz), which will need to be improved to be able to detect the typical ~200 µm wind-blown grains on Titan. We note that the PWA was designed with technologies from the 1990's and that experiments were performed at room temperature. There are therefore avenues for improving the noise level with the use of new technologies and the functioning in cold environments. This setup is a good analogue for the EFIELD experiment on Titan even if the configuration is slightly different from the one of the numerical simulations described in the previous section. In particular, for experimental simplicity, there is no wind, just a vertical fall of the particles whose terminal velocity is reached within ~300 ms, i.e. during the first ~15 cm in their falling trajectory.

Two additional instruments were added to precisely characterize the experiment. A floating metallic plate linked to an electrometer (6517A from Keithley) is used to measure the electric charge of the particles after their fall (as used in Houghton et al., 2013). A high-speed camera (240 fps, 320x240 pixels) films the ball falls to estimate their falling speed and their relative position to the electrode. All the instruments are positioned into a 90 cm x 60 cm x 65 cm Faraday cage to be shielded from the ambient electrostatic environmental noise. A 100 cm$^2$ opening at the top enables the passage of the charged falling particles, that are dropped a few centimeters above the Faraday cage. The complete setup is illustrated in Figure 10.

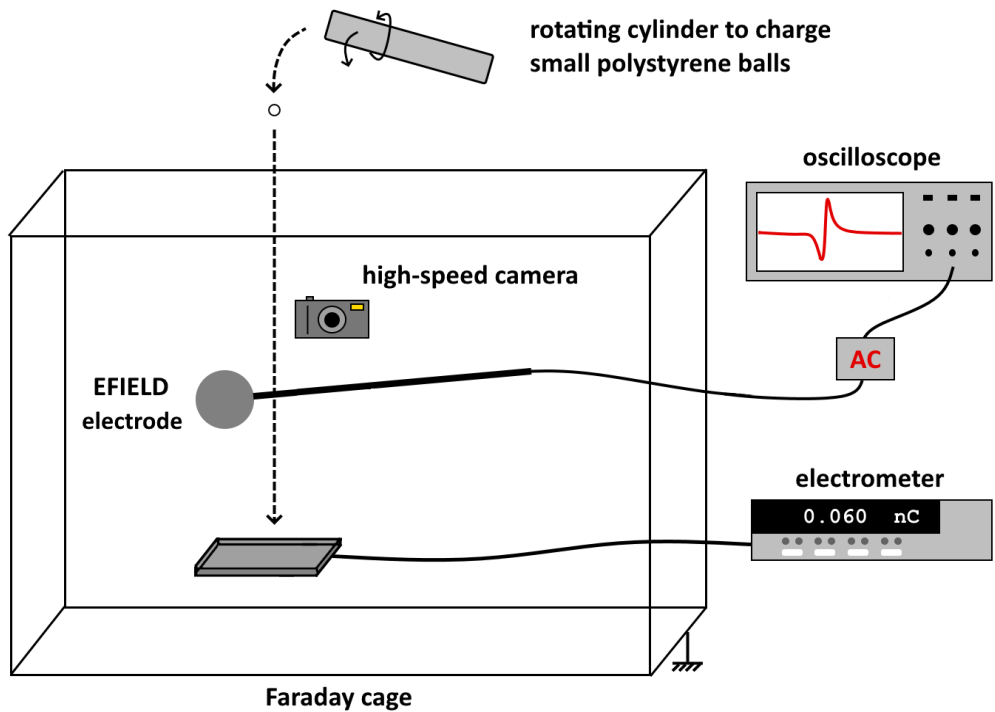

## 4.2- Validation of the inversion approach on experimental data

95 experiments were performed with set-up shown in Figure 10. In this experimental configuration, the assumption of linear trajectories applies perfectly and equation [3.2], combined to the measured transfer function (Figure S6-1 in Supporting Information), was used to fit the measured signals. In practice, this method works only if the input impedance of the electronic circuit is much larger that the impedance of the sphere. This is the case here. The input capacity of the receiver $C_R$ is estimated to ~0.3 pF (Hamelin and Chabassière, 2011). The capacity of the electrode $C_S$ can be estimated with the approximation of the spherical capacitor described further in the paper. Then, $C_S = \frac{Q}{U_{close}} = 4\pi\epsilon_0 \frac{R_S R_C}{R_C - R_S}$ with $R_C$ estimated to 30 cm (see below). It gives a capacity of the electrode of ~7 pF. Cables between the electrode and the receiver are bootstrapped and therefore do not participate to the attenuation of the signal, which is then given by the ratio of impedances $\frac{Z_S}{Z_S + Z_R}$, equivalent to $\frac{C_R}{C_R + C_S}$ in our case (see the equivalent electric scheme in section S8 in Supporting Information). The attenuation of the signal is then ~4%. This is a small effect that we did not take into account in this preliminary work. We also discarded the measurements associated with low signal-to-noise ratios to only consider 56 experiments among the 95; all done with balls of either 3 or 5 mm in size. Figure 11 displays some examples of measurements and their associated fits.

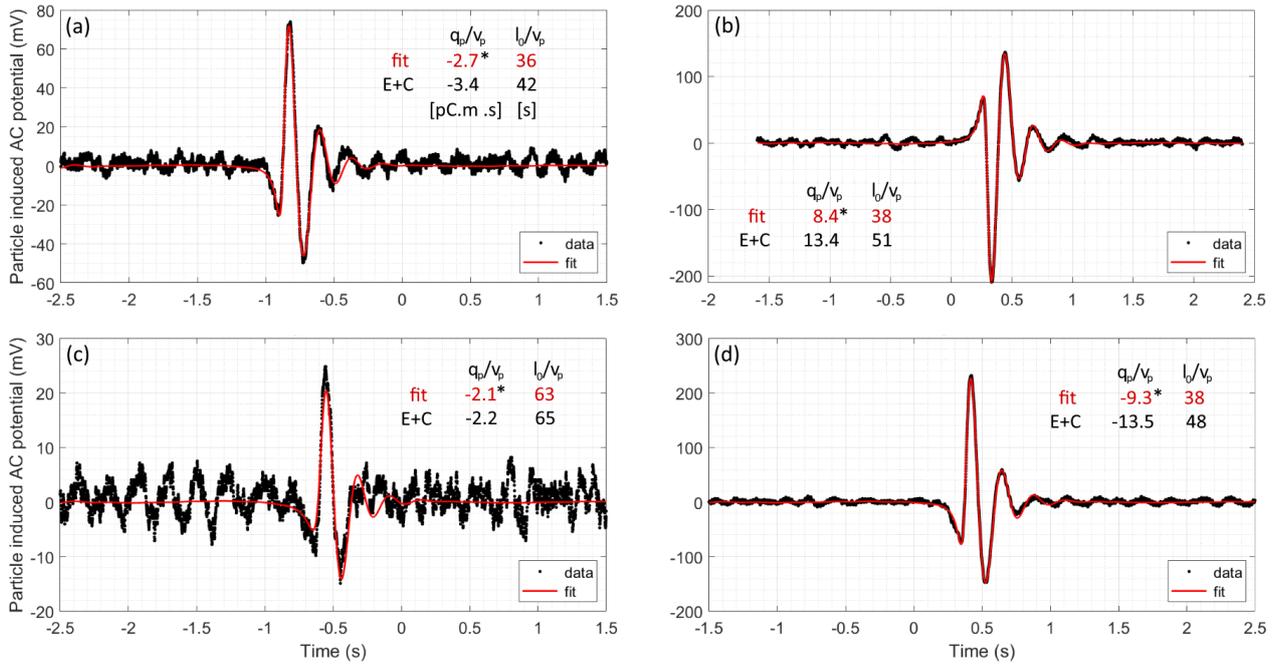

For all selected experiments, the values of $q_p/v_p$ and $l_0/v_p$ retrieved from the fitting of the data are compared to the values estimated with the electrometer and the high-speed camera. These two sets of values correlate very well as shown on Figure 12. The largest source of uncertainty comes from the camera measurements, which are limited by the fact that the camera is necessarily positioned close to the electrode in the Faraday cage, which affects its field of view. The projection of the camera image tends to underestimate the falling speed of the particle (see details in section S9 in Supporting Information). This is very likely the explanation of the shift coefficient observed in the scatterplot of $l_0/v_p$ on Figure 12b.

The offset between the expected and retrieved values of $q_p/v_p$ is more pronounced; though there is a clear correlation between these values, a systematic offset factor 0.71 is observed. Part of it can be explained by the underestimated particle falling speed as for $l_0/v_p$ (with the same coefficient 0.86). The other part (0.71/0.86 = 0.83) is most likely due to the presence of the Faraday cage around the experimental set up. The cage and the electrode act as a capacitor which reduces the measured potential by a constant whose value depends on the geometry of the system. The accurate estimation of the value of this constant would require a detailed model of the system which is out of scope of this paper. Nevertheless, a first order estimate can be obtained: a charge $Q$ on a conductive sphere of radius $R_S$ in open space corresponds to a potential $U_{open} = Q/(4\pi\epsilon_0 R_S)$ - if the sphere is itself surrounded by a larger conductive sphere of radius $R_C$ (proxy for the Faraday cage), the potential of the central sphere relatively to the outside sphere becomes $U_{close} = \frac{Q}{4\pi\epsilon_0} \times \left(\frac{1}{R_S} - \frac{1}{R_C}\right)$. This is the configuration of a spherical capacitor. The measured potential is thus decreased by a factor of $\left(1 - \frac{R_S}{R_C}\right)$ in presence of this simplified Faraday cage. An equivalent radius of 30 cm for the spherical Faraday cage would then explain the 0.83 missing coefficient since the value of $q_p/v_p$ is directly related to the amplitude of the potential (equation [3.2]). We note that this equivalent radius is a reasonable value since the dimensions of the rectangular cage are 90 cm x 60 cm x 65 cm.

In conclusion, the experiments performed with a preliminary prototype of the EFIELD electrode and the comparison of their results with the model we have developed to simulate the effect of a charged particle in the vicinity of the electrode provide further evidence that it will be possible to detect and characterize such particle with the EFIELD experiment. On Titan, the particle velocity will be estimated from wind measurements as well as camera observations and models. The retrieval of the $q_p/v_p$ parameters will thus be used to deduce particles charges, a key parameter to better understand sediment transport at the surface of the satellite. Further, on Titan, there will not be a Faraday cage; however, the presence of metallic surfaces in the vicinity of the EFIELD electrode (in particular that of the lander body of Dragonfly), will have an effect on the measured potential. Such effect can be readily measured and calibrated before launch. Lastly, we note that the hypothesis of a floating electrode may not apply depending on the final design of the electronic circuit of the EFIELD experiment; the equations will then have to be modified accordingly.

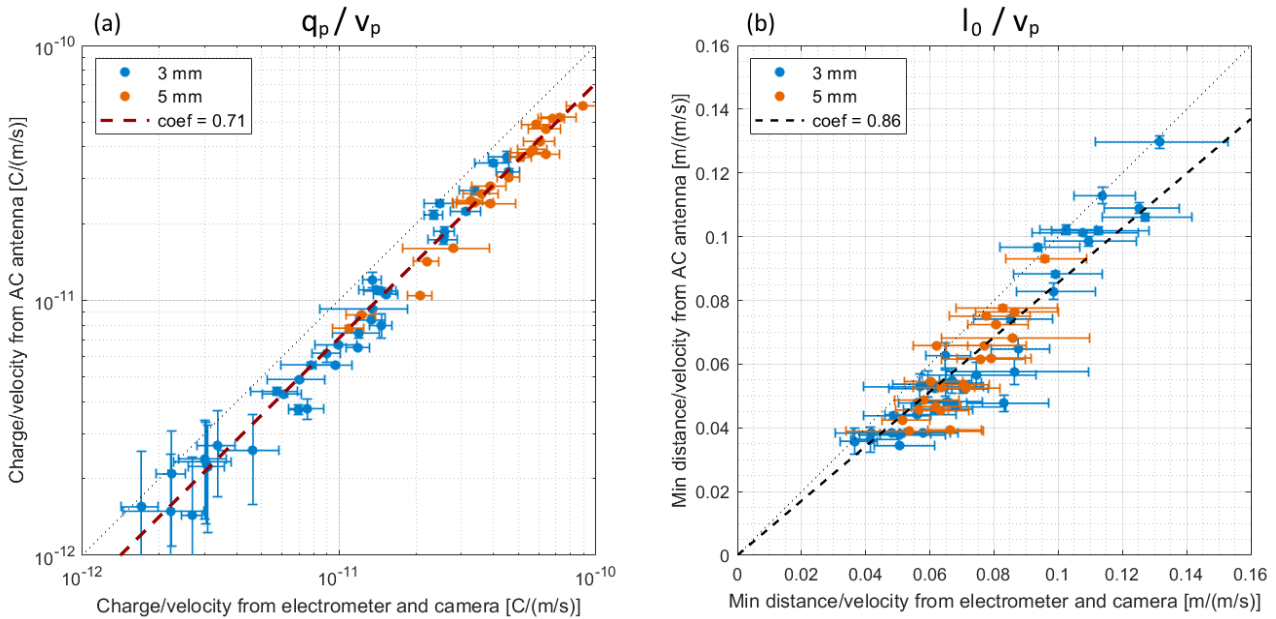

**Figure 12:** **Comparison of results with the AC electrode to measurements done with the electrometer and the high-speed camera. Statistics on 56 experiments with 3 and 5 mm polystyrene balls. AC electrode error bars are given by the 95% confidence interval of the fitting procedure. The light dotted line gives the perfect case; tentative explanations to its deviation are given in the text.**

## 5- Conclusion and perspectives

We have developed a numerical model to simulate the motion of a charged sand grain blown by winds close an idealized EFIELD electrode and the induced electric signal. This model assumes a floating electrode connected to an infinite input impedance preamplifier embedded in a neutral open space where the electrical conductivity can be neglected. We have then proposed an inversion method to derive the charge/velocity and the minimum distance/velocity ratios of the particle from its induced signal on the electrode. Lastly, we have set up a laboratory experiment to test a first prototype of the EFIELD electrodes and validate our numerical simulation and inversion approach. Perspectives for this work include tests in wind tunnels and field campaigns in snow-covered or sand-covered deserts. We note that this method is designed for the characterization of single particles: the fitting procedure works only on single-particle signals and gives the charge/velocity and the minimum distance/velocity ratios related to this particle. Individual grains cannot be resolved in a cloud of charged particles. The study of particle clouds could be the topic of a future complementary study, although large clouds of particles are not expected because of Titan's low surface winds and likely sticky surface.

Although the final design and accommodation of the EFIELD electrodes on the Dragonfly rotorcraft are not decided yet, this work suggests that charged sand grains blown by winds on Titan should, in many configurations, generate an electric signal strong enough to provide information on the particle characteristics and in particular, on their charge. The electronics of the EFIELD experiment has to be designed in order to provide a lower amplifier noise than the prototype electrode we used here. We recall that the standard deviation of the noise in our laboratory experiment (built with the 1990's technologies and tested in a warm environment) is $\sigma_n = 3$ mV. A reduction of $\sigma_n$ to a value of 1 mV would allow to detect particles of diameter 600 μm. Typical 200-μm grains could be observed close to the electrode for $\sigma_n = 0.5$ mV, though a value of 0.1 mV would be best for the analysis. We note that the density used to model particles was the one of fluffy analogues of Titan's aerosols ($\rho_p = 210$ kg/m$^3$, assuming a bulk density of 1400 kg/m$^3$ and a porosity of 85%, Brouet et al., 2016). However, sand grains on Titan are likely to have a lower porosity after years of transport at the surface (possibly 0 to 30%). In that case, grains would be denser. A charge-to-mass ratio identical to the one taken for the fluffy grains in this paper would lead to grains with higher charges, and therefore more easily detected. The uncertainty on the charge-to-mass ratio of such a material being of one order of magnitude or more, the accurate determination of the material density is not of major concern in this work.

One could wonder if –like for the Mars missions- the Dragonfly lander and the EFIELD antennas could be covered with fine organic grains ("dust") at some point during the mission. If this were to happen, the signal measured by the electrode would be attenuated. The measurement of a reproducible electric signal (e.g. from another instrument) all along the mission could then allow to monitor the coating of the electrode over time. Depending on their final orientations, cameras could also help to visualize the dust coverage. Nevertheless, several points will limit the dust accumulation on the electrodes. First, models and experiments predict negatively charged particles. Therefore, the electrode will certainly become negatively charged by collisions with a few initial particles, so that particles will then be repelled by the electrode. In addition, the ignition of the rotors and the regular flights through the dense atmosphere will also be efficient ways to remove dust deposited on the lander surfaces. And if this is still needed, the electrode could be brought to a high potential to detach all the remaining dust particles, as was done on the Langmuir probe electrode on-board Cassini after crossing Titan's ionosphere (Chatain et al., 2021).

We also recommend that at least one of the two EFIELD electrodes be located below the vehicle, at a few tens of centimeters from the ground, where particles in suspension or saltation are more likely to be found. We highlight that the signal generated by charged particles will not impede the main goal of the instrument which is the detection of Schumann resonances since – contrarily to single particle analysis – their detection requires an integration over a certain period of time in the frequency domain.

The measurement of the charge of wind-blown particles with the EFIELD experiment is key for the understanding of the aeolian activity on Titan. Such parameter is especially difficult to measure on a planetary body, and the method presented here proposes a unique way to perform such a measurement at the different landing sites of the Dragonfly lander, that is under a variety of wind and soil conditions, during at least the nominal duration of 3.3 years of the mission.

## Data availability

The laboratory measurements, the measured transfer function of the preamplifier, and the Matlab codes to analyze the data and plot the figures are available on Zenodo at https://doi.org/10.5281/zenodo.7274939. The Supporting Information document is also available at this link.

## Acknowledgements

ALG is supported by the Institut Universitaire de France (IUF).